\documentclass[superscriptaddress,twocolumn,showpacs,preprintnumbers,amsmath,amssymb, prx, reprint, longbibliography, showkeys]{revtex4-1}
\usepackage{amsmath}
\usepackage{amssymb}
\usepackage{graphicx}
\usepackage{stmaryrd}
\usepackage{txfonts}
\usepackage{mathrsfs}
\usepackage{dcolumn}
\usepackage{pbox}
\usepackage{bm}
\usepackage{epsfig}
\usepackage{color}

\usepackage[colorlinks=true, linkcolor=blue, citecolor=blue, urlcolor=blue]{hyperref}
\begin{document}

\title{Pairing symmetry and spontaneous vortex-antivortex lattice in superconducting twisted-bilayer graphene: Bogoliubov-de Gennes approach}
\author{Ying Su}
\email{yingsu@lanl.gov}
\author{Shi-Zeng Lin}
\email{szl@lanl.gov}
\affiliation{Theoretical Division, T-4 and CNLS, Los Alamos National Laboratory, Los Alamos, New Mexico 87545, USA}

\begin{abstract}
We study the superconducting pairing symmetry in twisted bilayer graphene by solving the Bogoliubov-de Gennes equation for all electrons in moir\'{e} supercells. 
With increasing the pairing potential, the system evolves from the mixed non-topological $d+id$ and $p+ip$ phase to the $s+p+d$ phase via the first order phase transition. In the time-reversal symmetry breaking  $d+id$ and $p+ip$ phase, vortex and antivortex lattices accompanying spontaneous supercurrent are induced by the twist. The superconducting order parameter is nonuniform in the moir\'{e} unit cell. Nevertheless, the superconducting gap in the local density of states is identical in the unit cell. The twist induced vortices and non-topological nature of the mixed $d+id$ and $p+ip$ phase are not captured by the existing effective models. Our results suggest the importance of long-range pairing interaction for effective models.
\end{abstract}

\date{\today}
\maketitle

\section{Introduction}
Pristine graphene is not superconducting. Surprisingly, by slightly twisting one layer in vertically aligned graphene bilayer, superconductivity emerges with a highest critical temperature $T_c\approx 1.7$ K \cite{cao_unconventional_2018}. The experimental twist angle is about $\theta=1.05^\circ$, which belongs to a discrete set of angles called magic angles. Theoretically, it was calculated that the electron velocity is quenched significantly at these angles \cite{bistritzer_moire_2011,PhysRevB.86.155449}, which has also been confirmed experimentally \cite{li_observation_2010,PhysRevLett.106.126802,PhysRevLett.109.126801,PhysRevLett.109.196802}. As a consequence, the kinetic energy of the electrons is reduced significantly and the role of electronic interaction becomes more important in the twisted bilayer graphene (TBLG). The misalignment between graphene layers induces a moir\'{e} superlattice (MSL) whose period is $a_M=a/2\sin(\theta/2)$ with $a=0.246$ nm the graphene lattice constant. At $\theta=1.05^\circ$, there exist four nearly flat bands around the Fermi energy according to an effective continuum model \cite{bistritzer_moire_2011,PhysRevB.86.155449}. At half filling of the lower two flat bands, a Mott-like insulating state has been observed in experiments \cite{cao_correlated_2018}. By doping the Mott-like state, superconducting domes appear. The phase diagram resembles that of high-$T_c$ cuprate superconductors, where the microscopic mechanism of superconductivity remains an open question, despite more than 30 years of effort. Unlike cuprates where chemical doping usually introduces unwanted side effects such as impurities, the electron concentration in TBLG can be controlled by using a gate voltage. Thus the TBLG offers a platform to investigate the mechanism of unconventional superconductivity.

The experiments have triggered tremendous theoretical endeavors to understand the origin of the Mott-like and superconducting state in TBLG. Several effective low-energy models based on MSL have been proposed to describe the flat bands at the magic angles \cite{xu_topological_2018,yuan_model_2018,rademaker_charge-transfer_2018,kang_symmetry_2018,wu_hubbard_2018,po_origin_2018,dodaro_phases_2018,koshino_maximally-localized_2018,you_superconductivity_2018,guinea_electrostatic_2018}. Based on these effective models, a variety of pairing symmetries, including spin singlet $d+id$ wave \cite{guo_pairing_2018,liu_chiral_2018,huang_antiferromagnetically_2018,kennes_strong_2018}, spin singlet but valley triplet $d+id$ wave \cite{roy_unconventional_2018}, spin triplet but orbital singlet $d+id$ wave \cite{xu_topological_2018,fidrysiak_unconventional_2018},  intertwined spin singlet (triplet) superconductivity with charge (spin) density-wave orders \cite{isobe_superconductivity_2018}, nematic and orbital triplet spin singlet superconducting state \cite{dodaro_phases_2018}, and extended $s$ wave \cite{ray_wannier_2018}, have been proposed. It is argued that intervalley electron pairing with either chiral ($d+id$ mixed with $p - i p$) or helical form factor is the dominant instability driven by the intervalley fluctuations \cite{you_superconductivity_2018}. These $d+id$ or $p+ip$ superconductors are topological, and can host nontrivial edge states \cite{black-schaffer_chiral_2014,kallin_chiral_2016}. Using the effective continuum model, both phonon-mediated $s$ wave and $d+id$ wave are proposed \cite{peltonen_mean-field_2018,wu_theory_2018}. It is argued that superconductivity arises from melting (doping) a Wigner crystal realized in TBLG \cite{padhi_wigner_2018}.  It is suggested that an emergent Josephson lattice is realized because of the inhomogeneous local density of states (LDOS) in TBLG \cite{baskaran_theory_2018}. It is also proposed that the Kohn-Luttinger instability in TBLG leads to an effective attraction between electrons and gives rise to superconductivity \cite{gonzalez_kohn-luttinger_2018}.  The nature of the Mott-like state has also been discussed extensively \cite{xu_kekule_2018,rademaker_charge-transfer_2018,pizarro_nature_2018,kennes_strong_2018,ochi_possible_2018,thomson_triangular_2018,po_origin_2018,wu_emergent_2018}.

The size of a moir\'{e} unit cell (MUC) is about 156 nm$^2$ at $\theta\approx 1.05^\circ$, and the LDOS is nonuniform. In principle, nontrivial superconducting texture in a MUC can arise, which cannot be captured by an effective low-energy Hamiltonian with electrons hopping in the MSL. The big size of the MUC allows for experimental detection of these superconducting textures.  Furthermore, an effective low-energy model for describing electrons in the MSL has not been firmly established yet \cite{po_origin_2018,zou_band_2018}. Mapping of the electron interactions from the original tight-binding model to the effective model is highly nontrivial. In this paper, we solve the Bogoliubov-de Gennes (BdG) equation by including all carbon sites in a MUC and by assuming a nearest neighbor (NN) pairing potential between electrons. For a weak pairing potential, a topologically trivial but time-reversal symmetry (TRS) breaking $d+id$ mixed with $p+ip$ spin singlet pairing state is stabilized. At a strong pairing potential, the system is stabilized at a nematic time-reversal invariant $s+p+d$ pairing state. The transition between these two states is of the first order. In the mixed $d+id$ and $p+ip$ state, vortices and antivortices in the MUC accompanying a spontaneous supercurrent are induced by twist. The superconducting order parameter is maximal at the center of a MUC. Nevertheless, the superconducting gap in the LDOS is identical in the MUC.

The bilayer superconducting systems can be regarded as two band superconductors. In the two-band superconductors, there exist metastable phase soliton excitations \cite{PhysRevLett.88.017002}. The phase soliton can be generated by injecting electrical current in nonequilibrium superconducting wires or by the proximity effect \cite{PhysRevLett.109.227003}. Such a phase soliton has only been observed recently in bilayer thin films \cite{PhysRevLett.97.237002}. In this work, we uncover a new mechanism to generate phase soliton lattice in mechanically twisted bilayer superconducting wires when the TRS is broken, see Sec. \ref{vortices}. In twisted bilayer films, the twisted induced topological excitations become the vortex and antivortex lattices. We remark that vortices are fundamental topological excitation in superconductors that govern the physical properties of the superconductors. Metastable vortices can be induced by defects or thermally fluctuations. The known mechanism for creating a vortex lattice requires an external magnetic field.

The paper is organized as follows. In Sec. \ref{mm}, we detail our model and numerical methods of simulating the superconducting TBLG. Then we study possible pairing symmetries in Sec. \ref{ps}. A quantum phase transition between two distinct superconducting phases is identified. Moreover, we find vortex and antivortex lattices appearing in the superconducting TBLG  when the TRS is spontaneously broken. In Sec. \ref{vortices}, we use a toy model to illustrate how twist can induce vortices and antivortices without external magnetic field. Finally, we show the spontaneous supercurrent and local density of states in Secs. \ref{ss} and \ref{ldos}, respectively.

\section{Model and method}
\label{mm}

In experiments, the transition from normal state to superconducting state is of the Berezinskii-Kosterlitz-Thouless transition due to the strong fluctuation of global phase of the superconducting order parameter in two dimensions. As far as the pairing symmetry of the superconducting order parameter in a MUC is concerned, we can neglect the fluctuation of the global phase and adopt a mean-field approximation in the following. The superconductivity is suppressed by a magnetic field parallel to TBLG, which indicates a spin singlet pairing. Moreover, the similarity between the cuprate and TBLG superconducting phase diagrams motivates us to introduce a singlet superconducting pairing potential between NNs. Such a pairing interaction can originate from the mean field treatment of the resonant valence bond interaction in $t$-$J$ model \cite{black-schaffer_chiral_2014}. For a single layer graphene, functional renormalization group calculations have revealed the emergence of singlet pairing interaction between electrons in the nearest or next-nearest neighborhood \cite{PhysRevB.86.020507}.  A recent calculation \cite{wu_theory_2018} based on phonon mechanism has demonstrated the existence of the assumed pairing interaction in TBLG. The direct interlayer pairing interaction is neglected because sites are misaligned vertically in TBLG. We neglect the Coulomb interaction. With these simplifications, the Hamiltonian for TBLG can be written as 
\begin{equation}
    \mathcal{H}=\sum_{ij\sigma}t_{ij}c_{i\sigma}^\dagger c_{j\sigma} - V\sum_{\langle ij\rangle}h_{ij}^\dagger h_{ij} - \mu\sum_{i\sigma} c_{i\sigma}^\dagger c_{i\sigma},
    \label{H}
\end{equation}
where $c_{i\sigma}^\dagger$ $(c_{i\sigma})$ creates (annihilates) an electron at site $i$ with spin $\sigma=\uparrow$ or $\downarrow$,
and $\langle ij \rangle$ denotes the two NN sites. Both the intralayer and interlayer hopping of electrons in TBLG are encoded in the first term. The hopping energy $t_{ij}$ depends on the vector $\bm{r}_{ij}=(x_{ij},y_{ij},z_{ij})$ connecting the two sites as \cite{PhysRevB.82.121407,PhysRevB.87.205404}
\begin{equation}
    t_{ij} = t_0 \exp\left(-\beta\frac{r_{ij}-b}{b}\right)\left(\frac{x_{ij}^2+y_{ij}^2}{r_{ij}^2}\right) + t_1\exp\left(-\beta\frac{r_{ij}-d}{b}\right)\frac{z^2_{ij}}{r_{ij}^2},
\end{equation}
where $t_0=-2.7$ eV, $t_1=-0.11t_0$,  $\beta=7.2$ characterizes the decay of hopping energy, $b=a/\sqrt{3}$ is the distance between NN intralayer carbon atoms, and $d=0.335$ nm is the distance between the two graphene layers. Here the interlayer hopping is truncated at $r_{ij}\leq 4b$ and only the NN intralayer hopping is considered.
The second term in Eq. (\ref{H}) depicts the spin-singlet pairing interaction with $h_{ij}=\left(c_{i\downarrow} c_{j\uparrow} - c_{i\uparrow} c_{j\downarrow} \right)/\sqrt{2}$, and the last term is the chemical potential. The model for a doped single layer graphene has been studied where a topological $d+id$ superconducting state is stabilized \cite{black-schaffer_resonating_2007}.

\begin{figure}
  \begin{center}
  \includegraphics[width=8 cm]{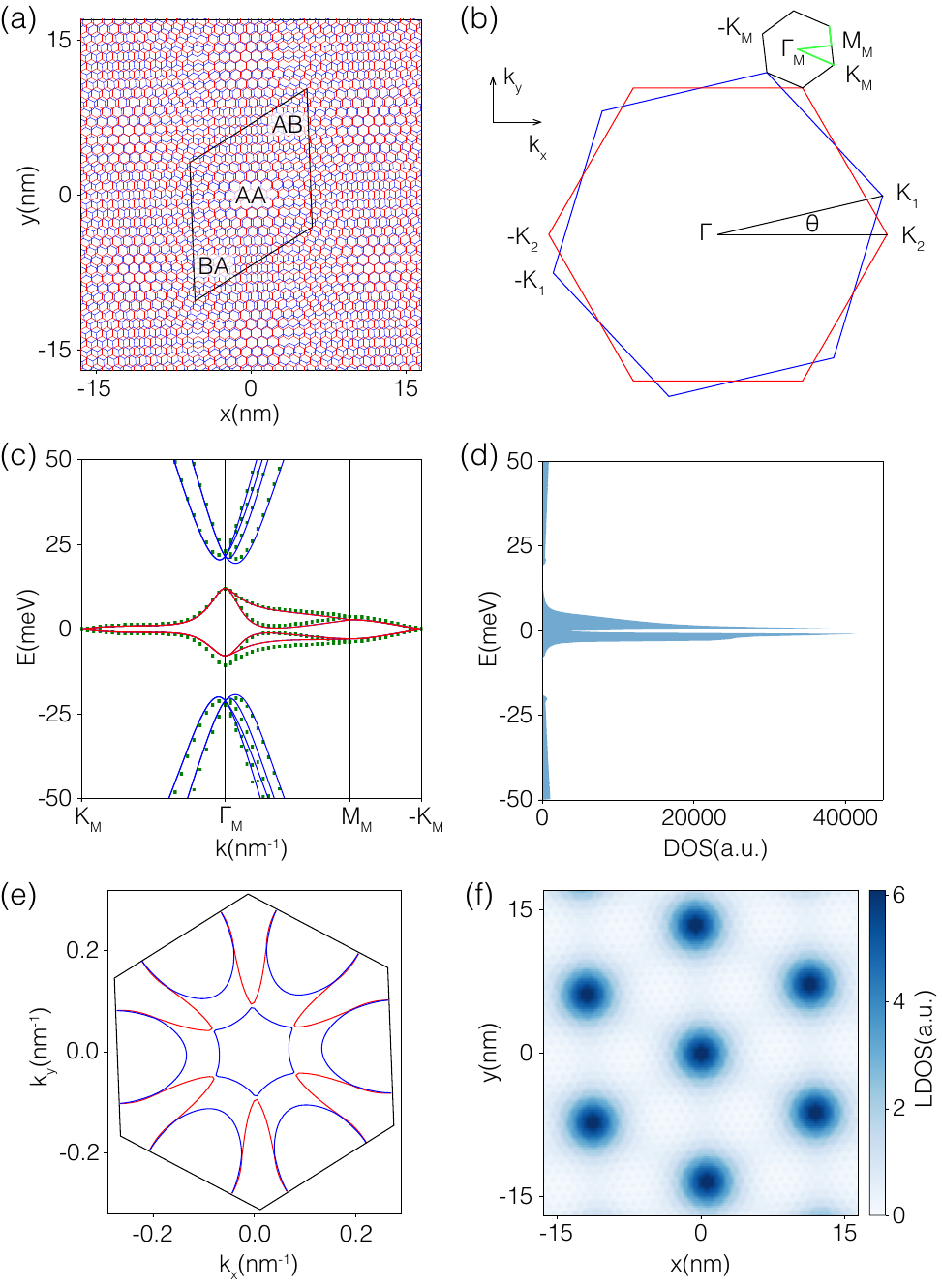}
  \end{center}
  \vspace{-0.2in}
\caption{\footnotesize{(color online). (a) Rescaled TBLG lattice. The black parallelogram encloses a MUC. AA, AB, and BA mark three different stacking patterns in a MUC. (b) The schematic BZs of the two graphene layers (red and blue hexagons) and the MBZ of the TBLG (black hexagon). (c) The band structure of the TBLG. The green points are from the unrescaled model, while the curves are from the rescaled model. The red curves denote the four flat bands.  (d) The DOS of the TBLG. (e), (f) Fermi surface and LDOS of the TBLG at the filling $\nu=(N-1.8)/N$, where $N$ is the total number of energy bands.}} 
  \label{fig1}
  \vspace{-0.2in}
\end{figure}

The periodicity and Fermi velocity of a TBLG are determined by the twist angle $\theta$. Strictly speaking, a TBLG has translation symmetry only at certain  discrete commensurate angles \cite{PhysRevB.81.165105}, 
\begin{equation}
\theta = \cos^{-1}\left(\frac{3p^2+3pq+q^2/2}{3p^2+3pq+q^2}\right),
\end{equation}
where $p$ and $q$ are coprime positive integers. For the other twist angles, the TBLG is not translation invariant though the period of MSL is still well defined. The lattice constant of commensurate TBLG is $a_C = qa_M/\sqrt{{\rm gcd}(q,3)}$ and $a_C=a_M$ only when $q=1$ \cite{zou_band_2018}.
For the magic angle $\theta = 1.05^{\circ}$, the commensurate TBLG with $p=31$ and $q=1$ has $N=4(3p^2+3pq+q^2)=11908$ carbon atoms in a MUC, which makes it time-consuming for numerical simulations. On the other hand, the effective continuum models 
fail to capture the internal structure of superconducting order parameter in a MUC. Therefore, it is essential to find a way to reduce the computation complexity while maintaining the essential features of the electronic structure in MSL. In this work, we adopt a rescaling approximation introduced in Ref. \onlinecite{PhysRevLett.119.107201} that enables us to use a larger twist angler $\theta'$ and smaller MUC to mimic the flat band signature of the magic-angle TBLG. In this approximation, the model parameters are renormalized as
\begin{equation}
t_0' = \frac{t_0}{\lambda},\quad V'=\frac{V}{\lambda},\quad a'=\lambda a,\quad  d'= \lambda d,\quad  \lambda = \frac{\sin\frac{\theta'}{2}}{\sin\frac{\theta}{2}},
\end{equation}
such that the Fermi velocity and MSL constant are invariant under the renormalization. Here primed quantities are the parameters for the rescaled model. In our numerical simulation, we choose $\theta' = 4.41^{\circ}$ (for $p=7$ and $q=1$) and the number of carbon atoms in a rescaled MUC is $N = 676$.  The rescaled TBLG is shown in Fig. \ref{fig1}(a). The graphene layer 2 (blue honeycomb lattice) is rotated anticlockwise with respect to layer 1 (red honeycomb lattice) by the rescaled twist angle $\theta'$.  The black parallelogram encloses a MUC that can be separated into AA, AB, and BA regions depending on the local stacking pattern. The schematic Brillouin zone (BZ) for the two graphene layers (red and blue hexagons) as well as the TBLG (black hexagon) is shown in Fig. \ref{fig1}(b). Here $\pm\mathbf{K}_{l=1,2}$ and $\pm\mathbf{K}_M$ denote the inequivalent valleys in the graphene BZs and in the moir\'{e}  Brillouin zone (MBZ). Moreover, in the MBZ	, the $\mathbf{K}_1$ and $-\mathbf{K}_2$ valleys coincide at $-\mathbf{K}_M$, while the $-\mathbf{K}_1$ and $\mathbf{K}_2$ valleys coincide at $\mathbf{K}_M$ for ${\rm gcd}(q,3)=1$ \cite{PhysRevB.81.165105}.

To validate the model, we calculate the single-particle band structure when $V=0$ for both the rescaled and unrescaled TBLGs. The calculated band structure at $\theta=1.05^\circ$ for the high symmetry path $\mathbf{K}_M$-$\mathbf{\Gamma}_M$-$\mathbf{M}_M$-$(-\mathbf{K}_M)$ in the MBZ [marked by the green path in Fig. \ref{fig1}(b)] is displayed in Fig. \ref{fig1}(c). Here the green points are from the unrescaled TBLG, while the curves are from the rescaled TBLG. Apparently, the rescaled model shows a good approximation to the unrescaled model. There are four flat bands with spin degeneracy near the Fermi energy in undoped systems and they are marked by the red curves in Fig. \ref{fig1}(c). These four bands touch at the $\mathbf{K}_M$ and $-\mathbf{K}_M$ of the MBZ, as a consequence of the spatial inversion, time-reversal and $C_3$ rotation symmetry of the TBLG \cite{po_origin_2018}.  The weak hybridization between the states near $\mathbf{K}_l$ and $-\mathbf{K}_{l'}$ of the original graphene BZ opens a small gap at $\mathbf{K}_M$ and $-\mathbf{K}_M$, which is manifested as a dip in density of states (DOS) at $E=0$, as depicted in Fig. \ref{fig1}(d). Such a hybridization is neglected in the continuum model approximation \cite{PhysRevLett.99.256802,bistritzer_moire_2011}. The shape of the four flat bands is in a good agreement with those obtained from continuum model. The present model has a slightly higher bandwidth about $20$ meV. Our model correctly yields large band gaps between the flat bands and other bands, which is underestimated by the continuum model, according to the experiments \cite{PhysRevLett.117.116804}.

\begin{figure}
  \begin{center}
  \includegraphics[width=8 cm]{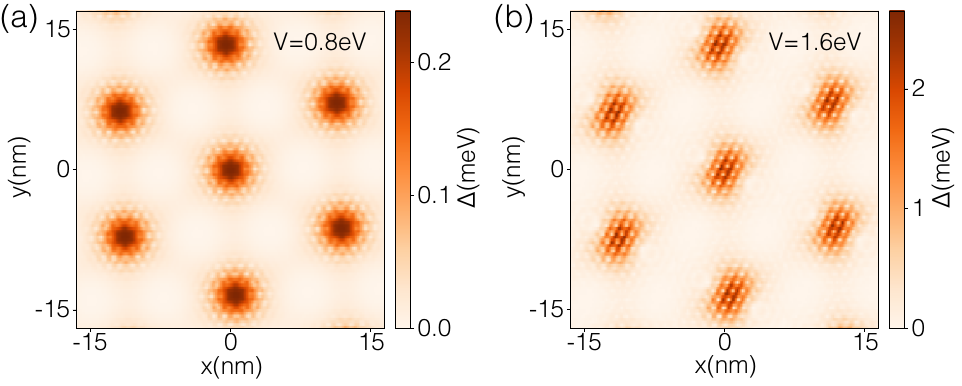}
  \end{center}
  \vspace{-0.2in}
\caption{\footnotesize{(color online). Superconducting order parameter distribution in the TBLG for the pairing strength $V=0.8$eV (a) and 1.6eV (b).}} 
  \label{fig2}
  \vspace{-0.2in}
\end{figure}

The four low-energy flat bands are mainly contributed from the four valleys of two graphene layers and each graphene layer provides two valleys. The four flat bands can at most host eight electrons in each MUC accounting for the spin degeneracy. The charge neutrality point corresponds  to the lower two flat bands, which are fully filled, while the upper two are fully empty. According to the experiments, there are Mott-like insulating phases at half fillings of the lower or upper two flat bands as a consequence of the reduced bandwidth \cite{cao_correlated_2018}. Moreover, in proximity to the half filling of the lower two flat bands, there are two narrow superconducting phases \cite{cao_unconventional_2018}.  Including the total $N$ (that is also the number of carbon atoms in a MUC) energy bands, the half filling of the lower two flat bands corresponds to the $(N-2)/N$ filling of the whole spectrum. Now we fix the filling at $\nu =(N-1.8)/N$, which corresponds to a slight electron doping of the half filling of the lower two flat bands, where superconductivity is observed in experiments \cite{cao_unconventional_2018}. The Fermi surface of the noninteracting magic-angle TBLG at the filling $\nu$ is shown in Fig. \ref{fig1}(e), where the red and blue curves respectively denote the Fermi surfaces from the lowest and second lowest flat bands. The Fermi surfaces are different from those of the continuum model due to the hybridization between the $\mathbf{K}_l$ and $-\mathbf{K}_{l'}$ valleys of the original graphene BZ, which is neglected in the continuum model.  The LDOS at the filling $\nu$ is shown in Fig. \ref{fig1}(f). The LDOS is inhomogeneously distributed in the MUC and the AA and AB (BA) regions correspond respectively to the region with maximal and minimal LDOS.  Because superconductivity is enhanced for a larger DOS. It is expected that spatially nonuniform superconductivity is stabilized in TBLG with a higher amplitude of superconducting order parameter in the AA region. This expectation is borne out by the numerical results shown below.

We now turn on the spin-singlet pairing in the magic-angle TBLG.  In the mean-field approximation, the pairing term becomes
\begin{equation}
 - V\sum_{\langle ij\rangle}h_{ij}^\dagger h_{ij}  = \sum_{\langle ij \rangle}\left[ \Delta_{ij}\left( c_{i\uparrow}^\dagger c_{j\downarrow}^\dagger - c_{i\downarrow}^\dagger c_{j\uparrow}^\dagger \right) + \text{H.c.}  \right],
\end{equation}
with $\Delta_{ij} = -\frac{V}{\sqrt{2}}\langle h_{ij} \rangle$.
We then solve numerically the corresponding BdG equation $\mathcal{H}_{\mathrm{BdG}}\Psi_m=E_m\Psi_m$ \cite{JXZhuBook}, where the BdG Hamiltonian is
\begin{align}
\mathcal{H}_{\mathrm{BdG}}=\nonumber\\
 \sum_{\langle ij\rangle} \left( {\begin{array}{*{20}{c}}
{c_{i \uparrow }^\dag }&{c_{j \uparrow }^\dag }&{{c_{i \downarrow }}}&{{c_{j \downarrow }}}
\end{array}} \right)\left( {\begin{array}{*{20}{c}}
{ - \mu }&{ - t_{ij}}&0&{{\Delta _{ij}}}\\
{ - t_{ij}}&{ - \mu }&{{\Delta _{ij}}}&0\\
0&{\Delta _{ij}^*}&\mu &t_{ij}\\
{\Delta _{ij}^*}&0&t_{ij}&\mu 
\end{array}} \right)\left( {\begin{array}{*{20}{c}}
{{c_{i \uparrow }}}\\
{{c_{j \uparrow }}}\\
{c_{i \downarrow }^\dag }\\
{c_{j \downarrow }^\dag }
\end{array}} \right).
\label{BdG}
\end{align}
The self-consistent equations for charge density $n_i$ and $\Delta_{ij}$ are
\begin{align}
n_i= \sum_m\left[ {\left| {u_{i \uparrow }^{m}} \right|^2}f\left( {{E_m}} \right) + {\left| {v_{i \downarrow }^{m}} \right|^2}f\left( { - {E_m}} \right)\right],
\end{align}
\begin{align}
\Delta_{ij}= \frac{V}{4} \sum_m \left[ {u_{i \uparrow }^m{{\left( {v_{j \downarrow }^m} \right)}^*} + {{\left( {v_{i \downarrow }^m} \right)}^*}u_{j \uparrow }^m} \right]\tanh \left( {\frac{{{E_m}}}{{2{k_B}T}}} \right),
\end{align}
where $\Psi_m=(\cdots,u_{i \uparrow }^m,\ u_{j \uparrow }^m, \ v_{i \downarrow }^m,\ v_{j \downarrow }^m,\cdots)^T$ and $E_m$ are the $m$-th eigenvector and eigenvalue of the BdG equation associated with Eq. (\ref{BdG}), and $f(E_m)$ is the Fermi function. 
In our numerical simulations, the TBLG contains $12\times12$ MUCs and has periodic boundary condition (PBC).
We solve the equations iteratively until $n_i$ and $\Delta_{i j}$ converge for all sites and bonds. We determine different superconducting phases by calculating $\Delta_{i j}$ at different $V$ and at zero temperature $T=0$.

 \begin{figure}
  \begin{center}
  \includegraphics[width=8 cm]{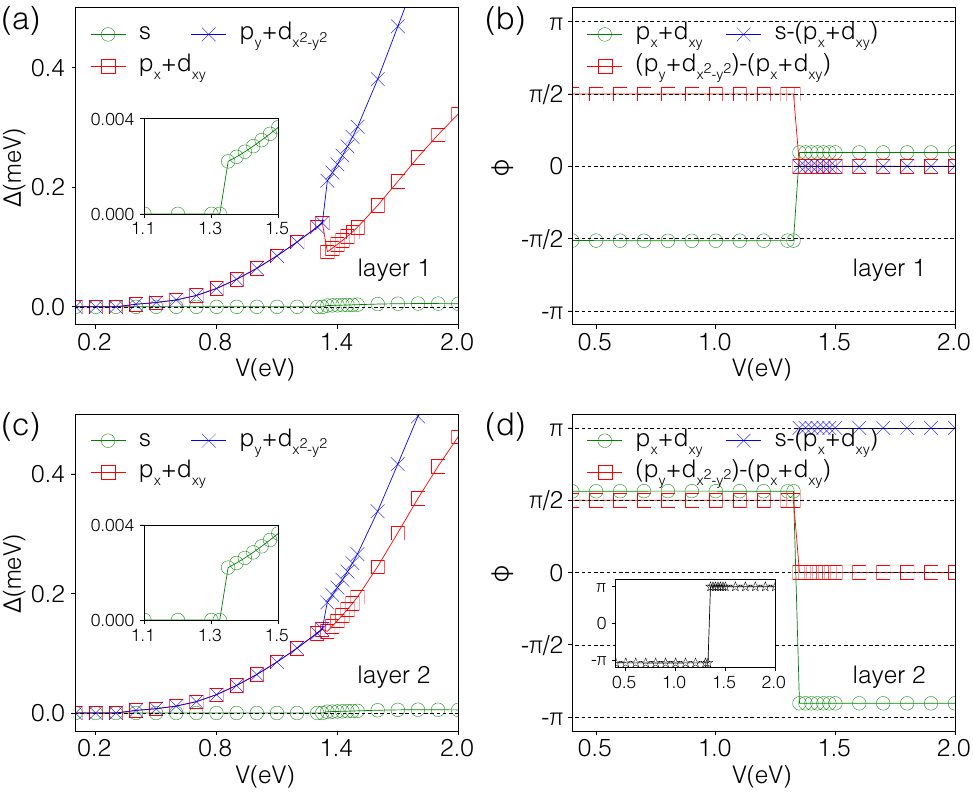}
  \end{center}
  \vspace{-0.2in}
\caption{\footnotesize{(color online). Averaged order parameter components $s$, $p_x+d_{xy}$, and $p_y+d_{x^2-y^2}$ in the two graphene layers. The amplitude of different components are shown in (a) and (c), and the phase difference between different components are shown in (b) and (d). The insets of (a) and (c) show the averaged $s$ component around the phase transition point. The inset of (d) shows the phase difference between the $p_x+d_{xy}$ components of the two layers. }} 
  \label{fig3}
  \vspace{-0.2in}
\end{figure}

\section{Pairing symmetry}
\label{ps}

\begin{figure*} 
  \begin{center}
  \includegraphics[width=17 cm]{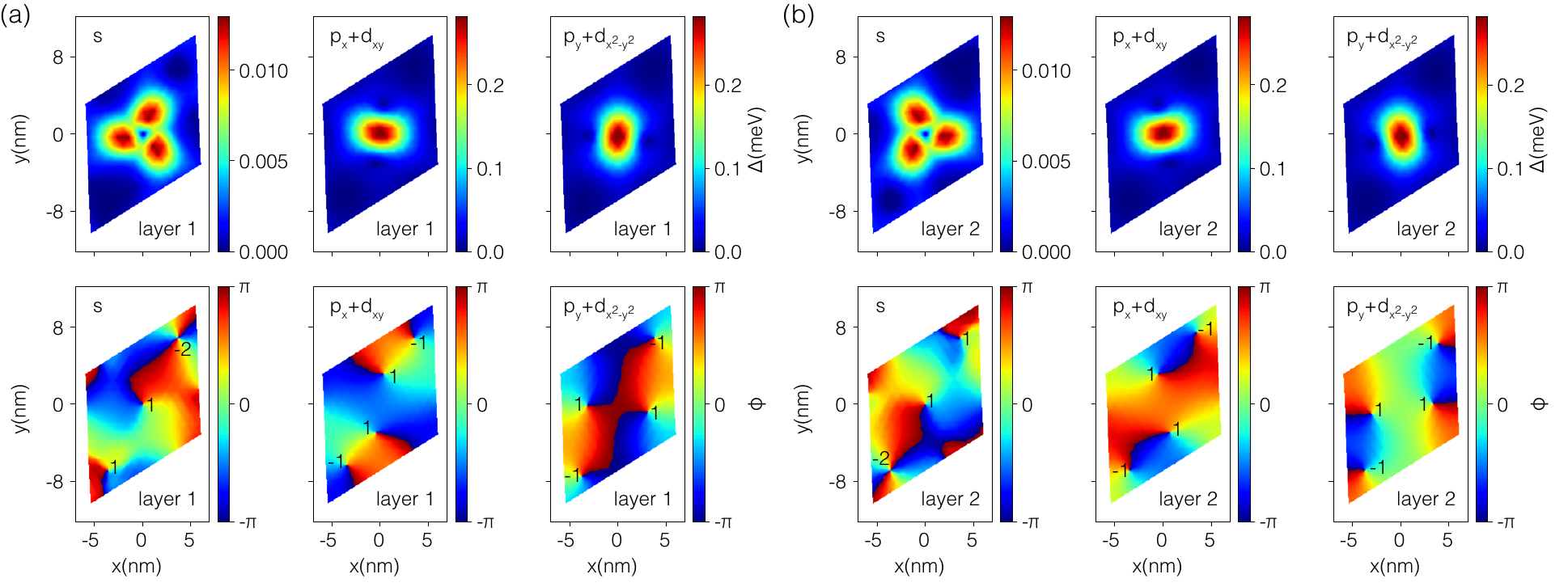}
  \end{center}
  \vspace{-0.2in}
\caption{\footnotesize{(color online). (a), (b) Profiles of different superconducting order parameter components for the two graphene layers in a MUC with $V=0.8$eV. The upper panels show the amplitude distribution and the lower panels show the phase distribution. The numbers in the lower panels denote the winding numbers of the vortices. }} 
  \label{fig4}
  \vspace{-0.2in}
\end{figure*}

\begin{figure*}
  \begin{center}
  \includegraphics[width=17 cm]{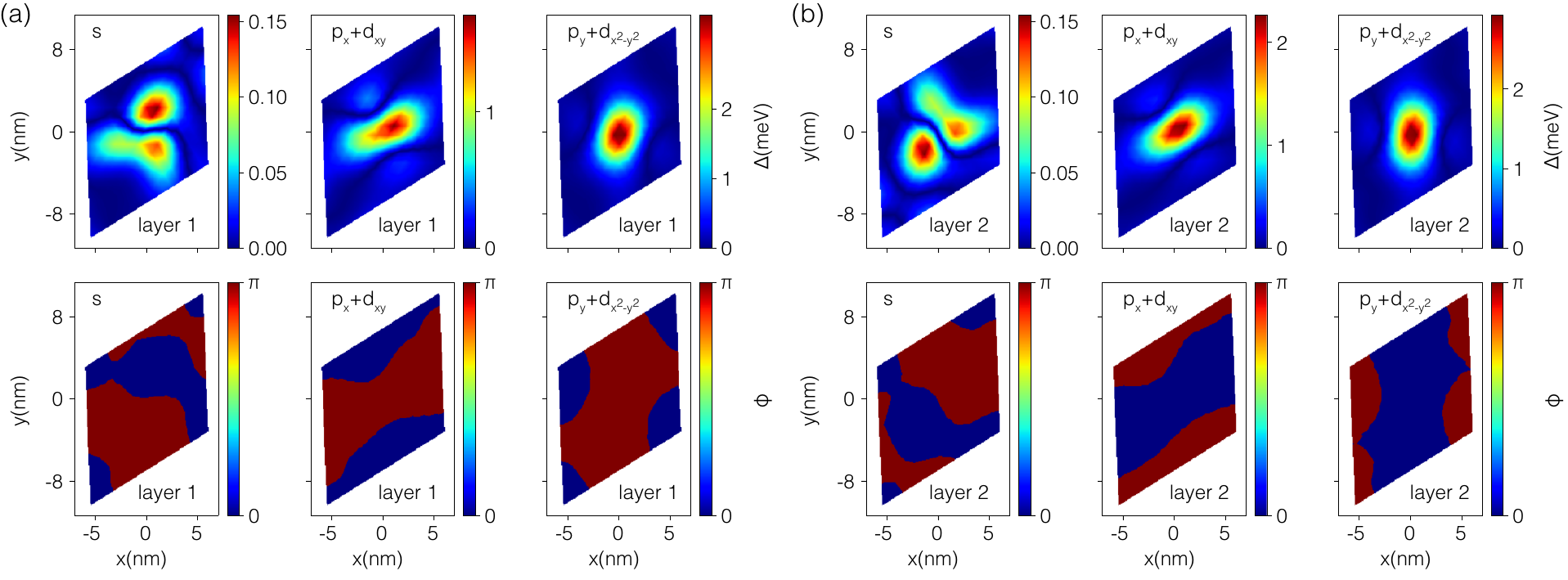}
  \end{center}
  \vspace{-0.2in}
\caption{\footnotesize{(color online). (a), (b) Profiles of different superconducting order parameter components for the two graphene layers in a MUC with $V=1.6$eV. The upper panels show the amplitude distribution and the lower panels show the phase distribution. }} 
  \label{fig5}
  \vspace{-0.2in}
\end{figure*}

In Fig. \ref{fig2}, we show the two typical profiles of the bond superconducting order parameter $\Delta_{ij}$ in the MSL for $V=0.8$ eV and $1.6$ eV. Due to the inhomogeneous electron concentration in each MUC, the order parameters are maximal in the AA region. Apparently, the order parameters preserve the rotation symmetry of the superlattice for $V=0.8$ eV; while for $V=1.6$ eV, the distribution of superconducting order parameter breaks the superlattice symmetry by elongating in the diagonal direction, and the resulting superconducting phase is nematic.

To elucidate the pairing symmetry and phase transition of the two distinct superconducting phases, we first consider the superconducting order parameter in different graphene layers. For graphene, it has $D_{6h}$ point group. We project the superconducting order parameter onto different irreducible representations of the $D_{6h}$ point group. For the $D_{6h}$ point group of graphene, the three dominant irreducible representations are $A_{1g}$, $E_{1u}$, and $E_{2g}$, which correspond to $s$ wave, $p$ wave, and $d$ wave, respectively.  Due to the $C_3$ rotation symmetry, the $p_y$ and $d_{x^2-y^2}$ components as well as the $p_x$ and $d_{xy}$ components  are  indistinguishable in the projection. Here we choose an orientation where  one graphene bond is in the $y$ direction. Because the superconducting order parameter is defined on the bonds connecting NNs, the basis functions for $s$ wave, $p_x$ and $d_{xy}$ waves, $p_y$ and $d_{x^2-y^2}$ waves are $\psi_1=(1,\ 1,\ 1)/\sqrt{3}$, $\psi_2=(0, -1, 1)/\sqrt{2}$, and $\psi_3=(2, -1, -1)/\sqrt{6}$, respectively. For a single layer graphene, the parity of orbital wave function in the superconducting wave function, i.e., $d$ wave or $p$ wave can be distinguished from the parity of the spin wave function. The pairing symmetry can be either $s$ or $d$ wave for the spin singlet or $p$ wave for the spin triplet to ensure the Fermi statistics for Cooper pairs. For TBLG, two bands originated from the single graphene $\mathbf{K}_l$ and $-\mathbf{K}_{l'}$ valleys are quasi-degenerate near the MBZ $\mathbf{K}_M$ and $-\mathbf{K}_M$ points, as discussed in Sec. \ref{mm}. Interband Cooper pairing between electrons in these two bands is also allowed. Here we call this pairing channel the intervalley pairing, which can also be decomposed into valley singlet or triplet in the valley degree of freedom. For the spin singlet pairing considered in Eq. (\ref{H}), we have $p$ wave valley singlet pairing and $d$ wave valley triplet pairing. {Because the symmetry of the TBLG is lowered compared to that of single layer graphene, and the nonlinear coupling of the superconducting order parameter is important at $T=0$, generally the superconducting order parameter of TBLG can be decomposed into a mixture of the different irreducible representations of the $D_{6h}$ point group.}

According to our calculations, $p$ wave and $d$ wave pairing coexist in TBLG. The phase difference between superconducting order parameters in different layers is about $\pm\pi$, as a result of layer counterflow velocity \cite{bistritzer_moire_2011}. In Fig. \ref{fig3}, we show the MUC averaged amplitude and phase of the $s$ component, mixed $p_y+d_{x^2-y^2}$ component, and mixed $p_x+d_{xy}$ component of the order parameters in the two graphene layers by projecting $\Delta_{ij}$ into $\psi_i$. As the pairing strength $V$ increases, the $p_y+d_{x^2-y^2}$ and $p_x+d_{xy}$ components firstly merge together and increase as displayed in Figs. \ref{fig3}(a) and \ref{fig3}(c). The amplitude of the superconducting order parameter grows as $\exp(-1/N_{\mathrm{DOS}} V )$ in accordance with a weak coupling theory. Our results differ from those obtained by the continuum model with an on-site attractive pairing interaction, where the amplitude of order parameter grows linearly with pairing interaction \cite{peltonen_mean-field_2018}. This discrepancy is due to the fact that the four low-energy bands are not completely flat, but are weakly dispersive with bandwidth about $20$ meV. Above a critical $V_c=1.325$ eV, the subdominant $s$ component jumps up from zero and the $p_y+d_{x^2-y^2}$ and $p_x+d_{xy}$ components separate. For $V<V_{c}$, the phase difference between the $p_y+d_{x^2-y^2}$ and $p_x+d_{xy}$ components is $\pi/2$ as shown in Figs. \ref{fig3}(b) and \ref{fig3}(d).  Namely, this superconducting phase is a mixture of $d+id$ and $p+ip$ in this region. The amplitude of the order parameter has $C_6$ rotation symmetry, but breaks the TRS. The states with phase difference $+\pi/2$ and $-\pi/2$ are degenerate and are related to each other through time-reversal transformation. For $V>V_{c}$, the phase difference between the $p_y+d_{x^2-y^2}$ and $p_x+d_{xy}$ components is $0$ for the two layers, while the phase difference between the $s$ and $p_x+d_{xy}$ components are $0$ and $\pi$ for the two layers. Therefore, we conclude the superconducting phase in this region is $s+p+d$, which breaks the $C_6$  rotation symmetry and is nematic.

\begin{figure}
  \begin{center}
  \includegraphics[width=8 cm]{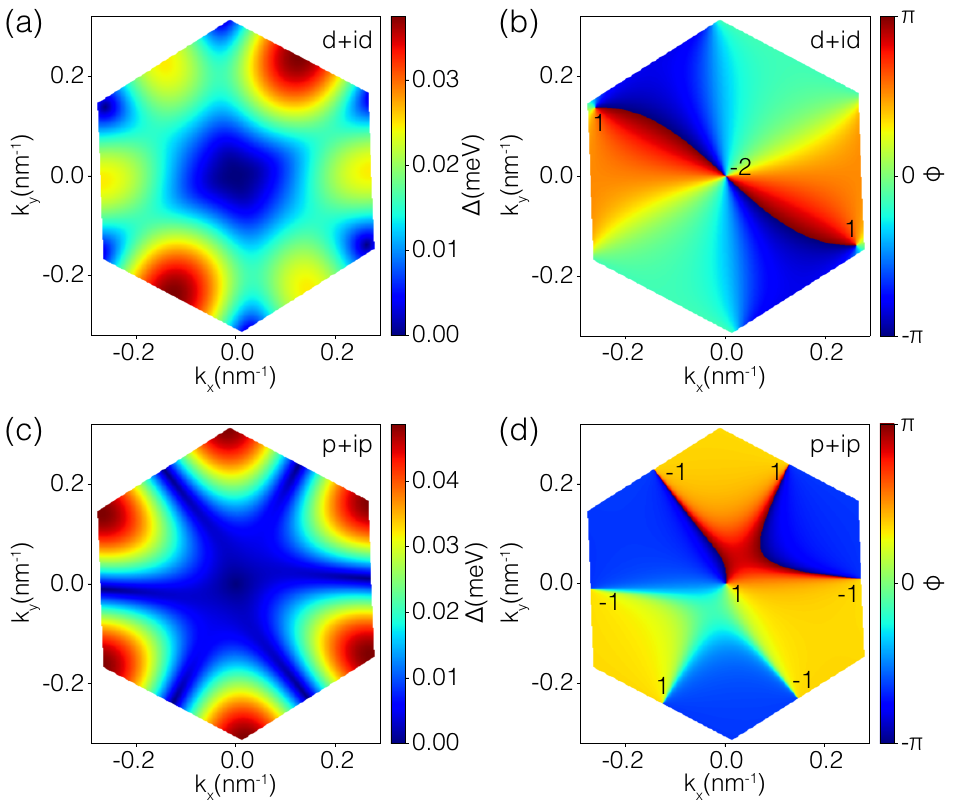}
  \end{center}
  \vspace{-0.2in}
\caption{\footnotesize{(color online). Profile of the superconducting order parameter $\Delta(k)$ in the MBZ with $V=0.8$ eV. (a), (b) Amplitude and phase of the even part of order parameter $\Delta_g(k)$. (c), (d) Amplitude and phase of the odd part of order parameter $\Delta_u(k)$. The numbers in (b) and (d) mark the winding numbers of the vortices. There are six vortices at the $\mathbf{M}_M$ points of the MBZ boundary in (d). These vortices are shared by two adjacent MBZs and contribute half winding number to the total vorticity in a MBZ which is zero.  }} 
  \label{fig6}
  \vspace{-0.2in}
\end{figure}

The amplitude and phase distributions of the three order parameter components for $V=0.8$ eV and $1.6$ eV in the two distinct superconducting phases are shown in Figs. \ref{fig4} and \ref{fig5}, respectively. Within one MUC, there are vortices associated with the winding of phases in all the pairing channels for $V=0.8$ eV. At the center of these vortices, the amplitude of the order parameter vanishes to avoid singularity, as shown in the upper panels of Figs. \ref{fig4}(a) and \ref{fig4}(b) for the two graphene layers, respectively. The total vorticity is zero in a MUC as shown in the lower panels of Figs. \ref{fig4}(a) and \ref{fig4}(b). The origin and consequence of these vortices will be discussed in Secs. \ref{vortices} and \ref{ss}. In contrast, there is no vortex in the $s+p+d$
 superconducting phase ($V=1.6$ eV). In stead, the order parameters are divided into different domains and the phase difference between different domains is $\pi$, as shown in Figs. \ref{fig5}(a) and \ref{fig5}(b).

\begin{figure}
  \begin{center}
  \includegraphics[width=8 cm]{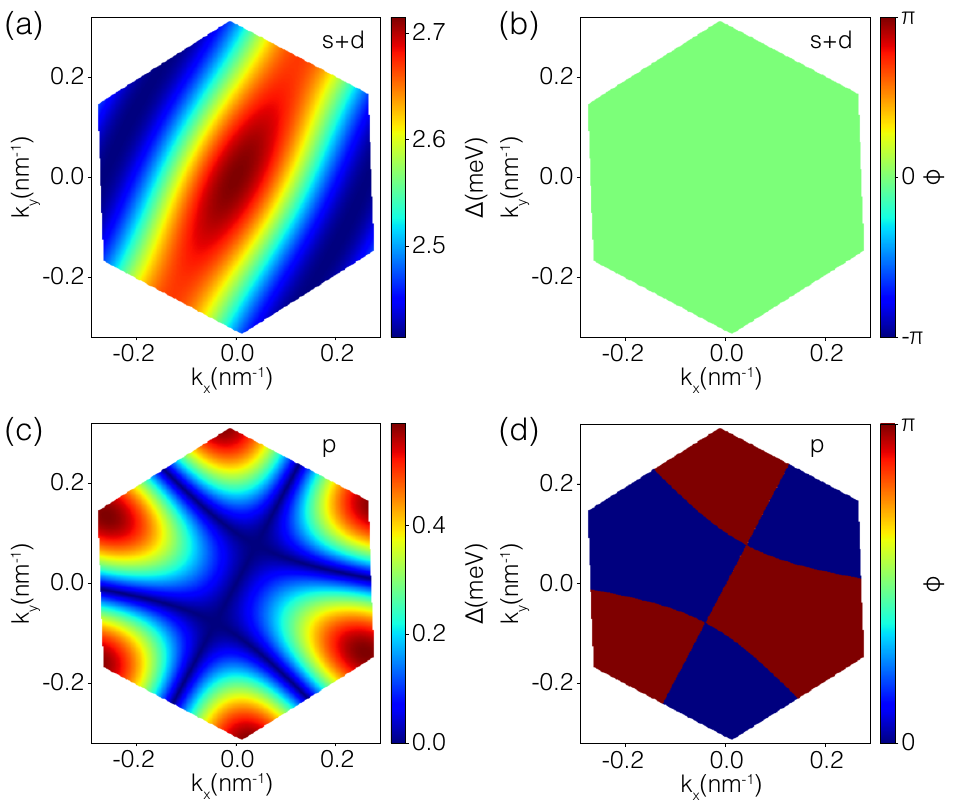}
  \end{center}
  \vspace{-0.2in}
\caption{\footnotesize{(color online). Profile of the superconducting order parameter $\Delta(k)$ in the MBZ with $V=1.6$ eV. (a), (b) Amplitude and phase of the even part of order parameter $\Delta_g(k)$. (c), (d) Amplitude and phase of the odd part of order parameter $\Delta_u(k)$.}} 
  \label{fig7}
  \vspace{-0.2in}
\end{figure}

To further distinguish the mixed $d$ and $p$ components, we perform the Fourier transform $\Delta(k)$ of order parameters $\Delta_{ij}$. The even part (for $s$ and $d$ components) of $\Delta(k)$, $\Delta_g(k)\equiv(\Delta(k)+\Delta(-k))/2$ and the odd part (for $p$ component), $\Delta_u(k)\equiv(\Delta(k)-\Delta(-k))/2$ in the MBZ are displayed in Figs. \ref{fig6} and \ref{fig7}. For $V=0.8$ eV, both the $d+id$ component [see Figs. \ref{fig6}(a) and \ref{fig6}(b)] and the $p+ip$  component [see Figs. \ref{fig6}(c) and \ref{fig6}(d)] show vortices at the MBZ center and with $4\pi$ and $2\pi$ phase winding, respectively. Besides the central vortices, there are additional vortices carrying opposite winding number and the net  vorticity in the MBZ is zero. Therefore, the mixed TRS breaking $d+id$ and $p+ip$ superconducting phase is topologically trivial. The amplitude of order parameter vanishes at the vortex cores. The quasiparticle spectrum is fully gapped because the Fermi surface does not cross the vortex cores. For $V=1.6$ eV, the dominant $s+d$ component [see Figs. \ref{fig7}(a) and \ref{fig7}(b)] has constant phase in the MBZ, while the subdominant $p$  component has $\pi$ phase difference between two domains [see Figs. \ref{fig7}(c) and \ref{fig7}(d)]. The superconductivity does not break TRS because $\Delta(k)$ and $\Delta^*(-k)$ are related by a global $U(1)$ transformation, i.e., $\Delta(k)=\Delta^*(-k)\exp(i\pi)$. 


\section{Twist induced vortices}
\label{vortices}

Our calculations reveal that vortices are nucleated by twist in the TRS breaking phase. To understand the origin of vortex at zero magnetic field, we consider a simple one-dimensional toy model. There are two chains with mismatched lattice constants $a_1$ and $a_2>a_1$. A MSL with period $a_M=a_1a_2/(a_2-a_1)$ is created by superimposing the two chains. Let us assume that each chain stabilizes a TRS breaking superconducting order parameter with $\Delta_{\alpha, j}=|\Delta_\alpha|\exp(\pm i j \phi_\alpha)$ when they are decoupled. Here $\alpha=1,\ 2$ labels the two chains and $j$ labels the bond, and $\pm$ corresponds to two degenerate states related by time-reversal operation. The interchain hopping of electrons yields a Josephson coupling between the two superconducting condensates, which favors the alignment of superconducting phase in different chains. An effective low-energy functional $\mathcal{F}$ for the superconducting phase $\phi_\alpha$ can be constructed,
\begin{align}
    \mathcal{F}=\sum_{\alpha=1,2}\frac{\eta_\alpha}{2}(\partial_x\phi_\alpha)^2 -J\cos(\phi_1-\phi_2),
\end{align}
subjected to the PBC that $\phi_\alpha$ changes by $\gamma_\alpha 2\pi$ in a MUC, with $\gamma_\alpha$ being an integer. Here $J$ is the interlayer Josephson coupling. Minimizing $\mathcal{F}$ with respect to $\phi_\alpha$, we obtain
\begin{align}
    \eta_\alpha\partial_x^2\phi_\alpha \mp J \sin(\phi_1-\phi_2)=0,
\end{align}
where $-\ (+)$ corresponds to $\alpha=1\ (2)$. The equation for $\phi_1-\phi_2$ is
\begin{align}
    \lambda_J^2 \partial_x^2(\phi_1-\phi_2)- \sin(\phi_1-\phi_2)=0,
    \label{fe}
\end{align}
with $\lambda_J\equiv \sqrt{\eta_1\eta_2/J(\eta_1+\eta_2)}$. For $a_M\gg \lambda_J$, Eq. \eqref{fe} allows for a soliton solution $\phi_1-\phi_2=\pm4\arctan[\exp(x/\lambda_J)]$. There are $\gamma_1-\gamma_2$ solitons of size $\lambda_J$ in a MUC. In this example, one can see that the TRS breaking allows $\phi_\alpha$ winds by multiple $2\pi$ in a MUC. The competing of Josephson coupling and this phase winding gives rise to phase slips with a sharp increase of phase by $2\pi$ in the MUC. 

The analysis can be generalized to the case of TBLG. In the TRS breaking superconducting phase in TBLG, the superconducting phase winds around the center of  a MUC. Similar to the chain chase, the competition between the phase winding and interlayer Josephson coupling induces phase slips in the MUC. The phase slips in two dimensions are vortices and antivortices, as we have observed in the BdG calculations. The generation of vortices requires to break TRS, but it does not require the superconductors to be topological.

\begin{figure}
  \begin{center}
  \includegraphics[width=\columnwidth]{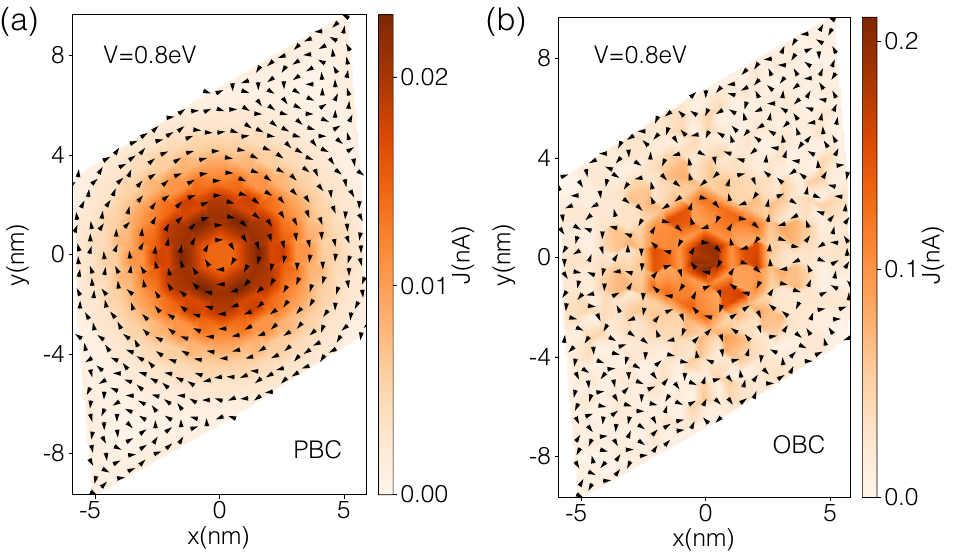}
  \end{center}
  \vspace{-0.2in}
\caption{\footnotesize{(color online). Distribution of spontaneous supercurrent for $V=0.8$ eV in a MUC with the PBC (a) and OBC (b), respectively. The black arrows denote the direction of the supercurrent and the background color represents the magnitude of the supercurrent.}} 
  \label{fig8}
  \vspace{-0.2in}
\end{figure}

\section{Spontaneous supercurrent }
\label{ss}

The twist-induced vortices result in spontaneous current and magnetization, which can be measured experimentally. In two-dimensional systems, the screening of magnetization by supercurrent is negligible, and we neglect it in our calculations. We calculate the current at each bond $\mathbf{J}_{ij}=-ie /\hbar \langle c_i^{\dagger} t_{ij} c_j-c_j^{\dagger} t_{ji} c_i\rangle\hat{e}_{ij}$, where $\hat{e}_{ij}$ is the unit vector pointing form site $j$ to site $i$. The spontaneous currents in the mixed $d+id$ and $p+ip$ state (for $V=0.8$ eV) are shown in Figs. \ref{fig8}(a) and \ref{fig8}(b) for the PBC and open boundary condition (OBC), respectively. For PBC, there are spontaneous current produced by vortices in both AA and AB (BA) regions of a MUC. Because the vortices carry opposite winding numbers (see Fig. \ref{fig4}), the current circulation directions are also opposite in different regions. The current is much stronger in the AA region because the superconducting order parameter is maximal there. For OBC, there are spontaneous supercurrent vortices in the AA region, while no current happens in the AB (BA) region due to the edge effect. The central current is stronger than that of the PBC. No edge supercurrent is observed, consistent with the nontopological nature of the $d+id$ ($p+ip$) phase discussed in the previous section. No spontaneous supercurrent is found in the $s+p+d$ phase.

\section{Local Density of States}
\label{ldos}

\begin{figure}
  \begin{center}
  \includegraphics[width=\columnwidth]{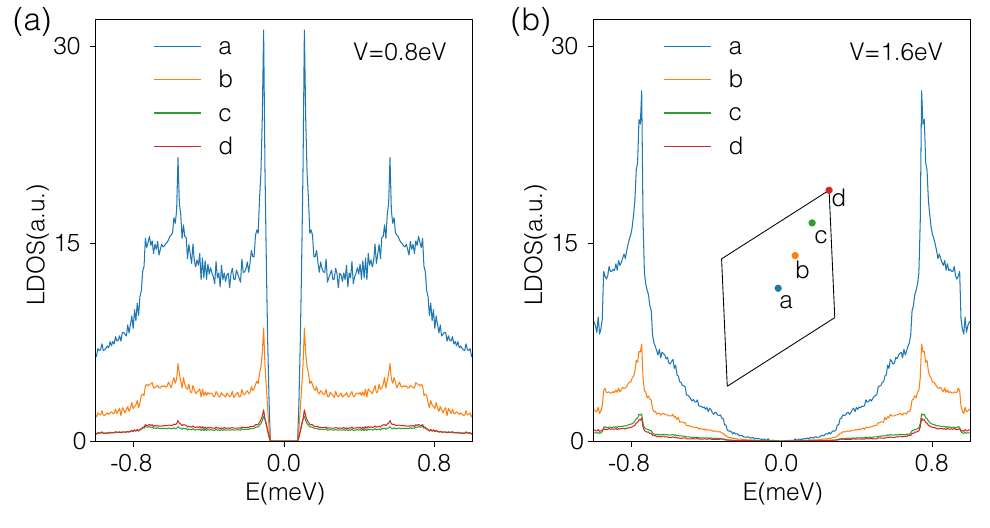}
  \end{center}
  \vspace{-0.2in}
\caption{\footnotesize{(color online). LDOS at four different points in a MUC for $V=0.8$ eV (a) and $1.6$ eV (b), respectively. The four different points marked as a, b, c, d are shown in the inset of (b), where the black parallelogram denotes the MUC.}} 
  \label{fig9}
  \vspace{-0.2in}
\end{figure}

The nonuniform superconducting order parameter in a MUC, at first sight, implies a variation of superconducting gap in the quasiparticle spectrum. We calculate the LDOS for both the mixed $d+id$ and $p+ip$ phase and $s+p+d$ phase at different sites in a MUC marked in the inset of Fig. \ref{fig9}(b). For both $V=0.8$ eV and $1.6$ eV, the LDOS is highest in the AA region and lowest in the AB (BA) region, consistent with those in Fig. \ref{fig1}(f). The quasiparticle spectrum is fully gapped at $V=0.8$ eV with the same gap at a different location even though the superconducting order parameter is small in AB (BA) region, as shown in Fig \ref{fig9}(a). The identical superconducting gap in the MUC can be understood as follows. First, the Fermi wave length of the moir\'{e} band is comparable to the MSL constant, although it is equal to 13.4 nm at $\theta=1.05^\circ$. In this sense, the distribution of superconducting order parameter in a MUC is in the ``atomic" limit.  Second, the scattering of quasiparticles in the AB (BA) region among strong superconducting AA region could produce an Andreev bound state. However, when the Fermi wave length is longer than the dimension of the weak superconducting region as in the present case, the quantum well produced by superconducting gap in the AA region is not deep enough to support any Andreev bound states. This can lead to identical gap with the gap determined by the superconducting gap in the AA region in a MUC. We note that the superconducting order parameters cannot be measured directly, because it is not gauge invariant. They are usually inferred from the quasiparticle spectrum.  Our results pose a question how the nonuniform superconductivity originated from the variation of LDOS in a MUC can be measured experimentally.    

At $V=1.6$ eV, the dominant pairing symmetry is $d$ wave. The subdominant $p$ wave has nodes at the Fermi surface.  The LDOS vanishes linearly as $|E|\rightarrow 0$, as shown in Fig. \ref{fig9}(b). The mixed $d+id$ and $p+ip$ pairing symmetry can be distinguished from the $s+p+d$ pairing symmetry  by measuring the LDOS.

\section{Summary and discussions}
\label{sd}

Our numerical results clearly reveal a textured superconducting order parameter with maximal amplitude in the AA region and minimal amplitude in the AB and BA regions. The textured superconductivity in a MUC was also proposed in Ref. \onlinecite {baskaran_theory_2018} based on the observation of nonuniform LDOS. The calculation based on the continuum model also results in a nonuniform superconductivity in a MUC both for the $s$ wave and $d$ wave superconductors \cite{peltonen_mean-field_2018,wu_theory_2018}. The Fermi wavelength for the moir\'{e} band is comparable to the moir\'{e} period, and the distribution of order parameter in a MUC can be regarded in the ``atomic'' limit. Despite that superconductivity is nonuniform in the length about $13.4$ nm for the magic twist angle $\theta=1.05^\circ$, the superconducting gap in the quasiparticle spectrum inferred from LDOS is identical in the MUC in the mixed $d+id$ and $p+ip$ superconducting phase. This poses questions how the textured superconductivity induced by variation of LDOS in a MUC can be measured experimentally.

We find two superconducting pairing symmetries, with a mixed $d+id$ and $p+ip$ spin singlet at weak pairing interaction and mixed $s+p+d$ spin singlet at strong pairing interaction, as shown in Table \ref{table}. A similar mixture of $d+id$ and $p-ip$ driven by valley fluctuations was discussed based on an effective model \cite{you_superconductivity_2018}. Nevertheless, the mixed $d+id$ and $p+ip$ phase breaks TRS, but it is topologically trivial, which is different from those considered in effective models. This discrepancy could be originated from the neglect of the long-range pairing interaction in these effective models. For the NN pairing interaction between MUCs, the form factors for the superconducting gap functions can only depend on $\cos k_x$, $\cos k_y$, $\sin k_x$, and $\sin k_y$. These form factors can describe the superconducting order parameter near the $\Gamma$ point of the MBZ. However, they are incapable of describing the antivortices away from the $\Gamma$ point in the MBZ. Our results suggest that long-range pairing interaction is important for the construction of an effective model.

\begin{table}
  \centering
  \vspace{0in}
\caption{\footnotesize{Summary of pairing symmetry (PS), pairing interaction strength (PIS), superconducting gap (SG), time-reversal symmetry (TRS), and topology of the two different superconducting phases.}}
\label{table}
\vspace{1em}
  \begin{ruledtabular}
  \begin{tabular}{ccccc}
  PS & PIS & SG & TRS & topology
  \\
  \hline 
    $d+id$ and $p+ip$  
  & $V<1.325$eV 
  & $\surd$
  & $\times$
  & $\times$ \\
    $s+p+d$ 
  & $V>1.325$eV   
  & $\times$
  & $\surd$
  & $\times$
  \end{tabular}
  \end{ruledtabular}
  \vspace{-0.1in}
\end{table}

We also reveal twist-induced vortice and antivortice lattice in the TRS breaking phase in TBLG. In the TRS breaking phase, there is winding of the phases of superconducting order parameter in different layers. Meanwhile, the interlayer Josephson coupling favors alignment of the phases of the superconducting order parameters. Because of the competition of these two effects, phase slips in the superconducting order parameters occur, which results in the nucleation of vortex and antivortex lattices. Spontaneous supercurrent and magnetization associated with the vortices and antivortices are generated, which can be detected by experiment using the standard magnetic imaging techniques. The mechanical twist provides a new mechanism to generate vortex and antivortex lattices in bilayer superconducting systems without the TRS.

To summarize, by solving the BdG equation for all electrons in a MUC, we show that the magic-angle TBLG with spin-singlet pairing interaction can support textured superconductivity due to the inhomogeneous distribution of electron density within a MUC. The two possible superconducting pairing symmetries are mixed $d+id$ and $p+ip$ wave and mixed $s+p+d$ wave. The mixed TRS breaking $d+id$ and $p+ip$ superconducting phase is topologically trivial. In the TRS breaking phase, the twist induces vortex and antivortex lattices in the MSL with spontaneous circulating supercurrent and magnetization. Though the superconducting order parameter is nonuniform in a MUC, the superconducting gap inferred from LDOS is uniform. In the $s+p+d$ state with TRS, no vortex is induced and the quasiparticle spectrum is not gapped.

\section{Acknowledgements}

Computer resources for numerical calculations were supported by the Institutional Computing Program at LANL. This work was carried out under the auspices of the U.S. DOE Award No. DE-AC52-06NA25396 through the LDRD program, and was supported by the Center for Nonlinear Studies at LANL and the U.S. DOE Office of Basic Energy Sciences Program E3B5 (S.-Z. L.).

\bibliography{references}

\end{document}